# A Protocol for Messaging to Extraterrestrial Intelligence

Dimitra Atri[1], Julia DeMarines[2] and Jacob Haqq-Misra[3]

[1]Department of Physics and Astronomy, University of Kansas (1082 Malott, 1251 Wescoe Hall Dr., Lawrence, KS 66045, USA, dimitra@ku.edu)

[2]Space Studies Program, International Space University (Parc d'Innovation,1 rue Jean Dominique Cassini 67400 Illkirch Graffenstaden. France, julia.demarines@masters.isunet.edu)

[3]Department of Meteorology, The Pennsylvania State University (407 Walker Building University Park, PA 16802, USA, misra@meteo.psu.edu)

**Corresponding Author**: Dimitra Atri, Department of Physics and Astronomy, University of Kansas, 1082 Malott, 1251 Wescoe Hall Dr., Lawrence, KS 66045-7550, USA.

Email: dimitra@ku.edu, Phone: +1-785-864-1272, Fax: +1-785-864- 5262

**Abstract**

Messaging to extraterrestrial intelligence (METI) is a branch of study concerned with constructing and broadcasting a message toward habitable planets. Since the Arecibo message of 1974, the handful of METI broadcasts have increased in content and complexity, but the lack of an established protocol has produced unorganized or cryptic messages that could be difficult to interpret. Here we outline the development of a self-consistent protocol for messaging to extraterrestrial intelligence that provides constraints and guidelines for the construction of a message in order to maximize the probability that the message effectively communicates. A METI protocol considers several factors including signal encoding, message length, information content, anthropocentrism, transmission method, and transmission periodicity. Once developed, the protocol will be released for testing on different human groups worldwide and across cultural boundaries. An effective message to extraterrestrials should at least be understandable by humans, and releasing the protocol for testing will allow us to improve the protocol and develop potential messages. Through an interactive website, users across the world will be able to create and exchange messages that follow the protocol in order to discover the types of messages better suited for cross-cultural communication. The development of a METI protocol will serve to improve the quality of messages to extraterrestrials, foster international collaboration, and extend astrobiology outreach to the public.

**1. Introduction**

The seminal publication by Cocconi and Morrison [1] suggests that if advanced extraterrestrial societies exist in the Milky way, then they may attempt interstellar communication across the galaxy. This idea led to the first search for extraterrestrial intelligence (SETI) by Frank Drake known as Project Ozma, which used a 26 meter radio telescope to examine two neighboring Sun-like stars near the 1.420 gigahertz frequency [2]. Nearly fifty years after Project Ozma, SETI has expanded to include searches for both radio and optical signals by *piggybacking* on conventional astronomical sky surveys. Additionally, dedicated SETI observing facilities, such as the Allen

Telescope Array, can target specific stellar systems to listen for an extraterrestrial transmission [3].

Many current planet detection methods are biased toward the discovery of gas giants, but extrasolar planet detection techniques will soon be able to detect habitable terrestrial planets through both ground and space based observations. The recently launched Kepler telescope will observe over 100,000 stars in the Milky Way in search of transiting Earth-sized planets, while missions such as the New Worlds Observer and the Terrestrial Planet Finder are being designed (and will hopefully have financial backing in the next decade) to directly detect spectroscopic signatures of extrasolar terrestrial planets [4]. Although the presence of life cannot be confirmed with these methods, a planet could be deemed a likely candidate for life based on its atmospheric composition and orbital position. For example, the simultaneous presence of oxygen and methane in a planet's atmosphere could be a strong indicator of biological processes [5]. Spectroscopic biosignatures can only provide evidence for the probable existence of life, but potentially inhabited terrestrial planets are also excellent targets for communication with extraterrestrials. For practical purposes, SETI considers intelligent life to be any society capable of radio transmission so that interstellar communication is possible. (According to this definition, Earth has had intelligent life for less than 100 years.) An Earth-like planet that shows spectroscopic indications of biological processes has a higher probability of housing intelligent life that has developed communicative technology. If (or perhaps, when) we find planets like this, they will be the best known targets for sending a message to extraterrestrials.

Communication with extraterrestrial intelligence has been widely discussed [6], most notably in the 1970's [7,8], but only a handful of attempts at messaging to extraterrestrials (METI) have been undertaken. The content of past METI broadcasts has also contained information dependent on human culture and biology, which may not be universally understood. Many people would choose to construct a message containing sights and sounds of the human experience [9], but overly anthropocentric signals that implicitly rely upon certain facets of human culture may go unnoticed by extraterrestrial listeners. METI may therefore increase its probability of success by decreasing the cultural dependence of messages.

Some critics of METI argue that broadcasting our presence to extraterrestrials is a security risk because it would alert a malevolent extraterrestrial civilization of our presence [10]. An extraterrestrial civilization that detects our transmissions will likely have superior technology to our own, so they would almost certainly posses the capability to destroy us. However, Earth has been emitting electromagnetic signals to space since the beginning of the radio age, mostly as unintended *leakage* from television, aviation, and telecommunication [11]. An advanced civilization within a radius of 100 light years could detect our television shows and already know we are here, so there is little hope in concealing our location in space. Extraterrestrials may react to an intentional or unintentional human broadcast with war, benevolence, apathy, or suicide, depending on their ethical framework [12], and total annihilation of one civilization by the other is not necessarily the most likely outcome. Additionally, we have not yet observed any extraterrestrial civilizations or received any extraterrestrial broadcasts, so this conspicuous absence remains to be explained [13,14]. Maybe complex life is rare [15], or maybe we are actually being stealthily observed [16,17]. But perhaps civilizations that rapidly expand across the galaxy are also quickly forced into ecological collapse, while less expansive civilizations that grow within their carrying capacity may not have had enough time to colonize the galaxy [18,19]. If this is the case, then we are unlikely to encounter an expansive extraterrestrial civilization, and those that we do encounter may be less inclined toward violent conflict.

Here we discuss the development of a protocol for METI to address the shortcomings of past broadcast attempts. A protocol for METI will provide guidelines for the message length, signal encoding, transmission method, and information content in order to maximize the likelihood that the message is understood. Successful messages will minimize any anthropocentric or culturally dependent content and avoid dependence on basic human senses. Additionally, a message that is expected to be decoded and comprehended by extraterrestrials should be decipherable across human cultural boundaries on Earth, so the construction of a METI protocol will provide a means to develop and test messages before they are broadcast.

**2. Previous messages to extraterrestrials**

The first and most well-known message was the Arecibo message sent from the Arecibo radio telescope in Puerto Rico via frequency modulated radio waves in 1974 [20,21]. It was targeted at the globular star cluster M13, which is approximately 25,000 light years away. The message was written by Frank Drake, with help from Carl Sagan and others, and consisted of 1679 binary digits (approximately 210 bytes) transmitted at a frequency of 2380 MHz and modulated by shifting the frequency by 10 Hz, with a power of 1000 kW [22]. (By comparison, strong television transmitters have a power of about 170 kW [11]. The "ones" and "zeroes" composing the message were transmitted by frequency shifting at the rate of 10 bits per second. The message consisted of seven parts that encode the following:

1. The numbers one through ten
2. The atomic numbers of the five elements hydrogen (H), carbon (C), nitrogen (N), oxygen (O), and phosphorus (P), which make up deoxyribonucleic acid (DNA)
3. The formulae for the sugars and bases in the nucleotides of DNA
4. The number of nucleotides in DNA, and a graphic of the double helix structure of DNA
5. A graphic figure of a human, the physical height of an average man, and the human population of Earth
6. A graphic of the Solar System
7. A graphic of the Arecibo radio telescope and the physical diameter of the transmitting antenna dish

Because it will take 25,000 years for the transmission to reach its destination and at least another 25,000 years to receive a reply, the Arecibo message was more of a demonstration of human technological achievements than a real attempt to communicate with aliens.

The second realized project was called Cosmic Call [23], which consisted of two interstellar radio messages that were sent from Evpatoria, Ukraine in 1999 and 2003 to various nearby stars. The formatting and characters used in the messages were designed to resist alteration by noise. The third, called "Teen Age Message", was transmitted from the Evpatoria Planetary Radar to six nearby Sun-like stars during August-September 2001 [23]. Unlike the previous digital-only messages, the Teen Age Message has a three-section structure containing various forms of information, suggested by Russian astronomer Aleksandr Zaitsev: Section 1 represents a coherent sounding radio signal with slow Doppler wavelength tuning to imitate transmission from Sun's center; Section 2 uses analog information and contains musical melodies performed on the theremin. This electric musical instrument produces a quasi-monochromatic signal, which is detectable across interstellar distances; Section 3 contains binary digital information including the Teen Age Message logo and bilingual Russian and English greetings to the aliens. The message is named after the Russian teens who arranged and performed the classical music compositions in Section 2.

The fourth and latest attempt was made in 2008 called "A Message From Earth" (AMFE), which was a digital radio signal sent towards Gliese 581c, a large terrestrial extrasolar planet orbiting the red dwarf star Gliese 581. This message was also emitted from the radar telescope at Evpatoria, Ukraine. Five hundred and one text messages, photographs and drawings submitted by the public were selected to be transmitted in the digital time capsule. These messages consisted of various topics, such as the submitters' own lives and ambitions, claims of world peace, and views of the Earth. These three recent broadcasts have targeted stars between 20 and 69 light years from Earth.

The description of past transmissions illustrates that the complexity and the anthropocentric nature of transmitted messages has increased significantly with time. The sophistication of human technology has enabled us to transmit more complex messages that include digital imagery, digital sound, and analog sound. However, greater complexity probably makes it more difficult for an extraterrestrial listener to decode and decipher the message. Given that we know very little about the nature of extraterrestrial civilizations, if they exist, we are likely to increase the probability of us successfully communicating to them if we use a message that the recipient is likely to understand. Messages that relate strongly to certain specifics of humankind or to one of our physical senses may be difficult to understand by a listener far removed from the contexts of Earth. Messages using proprietary compression or conversion algorithms also may be difficult for extraterrestrials to decode. The likelihood of success for METI may thus be improved by the construction of a protocol for messaging to extraterrestrial intelligence.

**3. Developing a Protocol for Messaging to Extraterrestrials**

Sending a message to an unknown extraterrestrial intelligence may require a loosening of the anthropocentric assumptions that have characterized previous METI attempts. Most communication between humans on Earth relies on vision and sound because of the specifics of our biology, but we cannot necessarily assume that extraterrestrials will share any of our basic senses. The dependence on visual imagery in the Arecibo message, for example, necessitates that the recipients can interpret information stored as pictures. However, a subsurface life form is less likely to develop visual sensory perception, and not even all life on Earth is vision-oriented (e.g., bats) or auditory (e.g., invertebrates). Likewise, while we expect that the basic physics of compression waves applies throughout the universe, not all life forms will necessarily develop a sense of hearing. Messages that rely on the specifics of human biology or culture will be less likely to effectively communicate to an unfamiliar extraterrestrial listener.

Additionally, the information content of METI signals has increased since the initial Arecibo message. Modern technology allows for large amounts of data to be transmitted at moderate costs, but the broadcast of massive amounts of information assumes that the recipient extraterrestrials will be capable of comprehending a complex message. Even if advanced extraterrestrial technology can decode a massive message with ease, a society of blind extraterrestrials would be unsuccessful at retrieving the information from a visual message just as deaf recipients would be unable to understand an auditory message. Extraterrestrials may identify a signal containing video or music as originating from Earth-based life because its complexity resembles nothing else found in nature, but they may have little to no success at comprehending the information within an overly complex message. Thus, it is imperative that any message intending to communicate with extraterrestrials be short and simple enough that it can be understood by the widest possible audience.

In order to improve upon past METI attempts, we propose to develop a METI protocol. This protocol will provide constraints and guidelines for the construction of a message in order to maximize the probability that the message is understood. A METI protocol will consider several factors, including:

1. Signal encoding
2. Message length
3. Information content
4. Anthropocentrism
5. Transmission method
6. Transmission periodicity

A transmitted message should be encoded in a way that can be recognized and interpreted by a broad audience. Past choices for encoding have included the use of a binary signal—ones and zeros—though frequency modulation allows for the use of an arbitrary base system. Certain wavelengths may be more common for communication than others, and repetition of a broadcast message toward a given target is almost certainly necessary. Complex and proprietary encoding (such as compressed audio) should be avoided, and an appropriately encoded message should be of a limited length so that the amount of information in the message is not overwhelming. A restricted message length will also be more cost effective in terms of power requirements. Past METI attempts have tended to increase in content, but without a proper cultural context much of this information may not be understood by an extraterrestrial listener. Limiting the extent to which the information depends on human-like traits will help guide the construction of a message. This may also require that the message attempts to represent Earth as a whole instead of focusing exclusively on humanity.

The construction of a consistent METI protocol (or several protocols) is a daunting task that requires a wide representation spanning a range of disciplines and extending further into the public sphere. Creation of an initial protocol would likely benefit from the formation of a METI committee that thoroughly investigates the factors involved in messaging and comes to a protocol by consensus.

**4. Transmission Strategy**.

Since it is almost impossible to decode a message without a knowledge of the language at the other end, we should rely on a simple physical or mathematical language to communicate both the encoding scheme and the content [24]. It is also important to regularly repeat the transmission because there are many ways for a detector to generate a false signal. Repetition of a broadcast allows the receiver to be certain that they have detected a genuine signal instead of machine noise. Periodic transmission will increase the probability of a signal being received because the receiving antenna may not always be tuned to the particular frequency and direction when the signal arrives. It may also be wise to repeat the fundamentals of the encoding language (e.g., the definition of addition) at a greater frequency than the more complicated parts of the message. The length of the message should be optimized to the number of targets selected and the periodicity of transmission.

There are several aspects that need to be taken into account when choosing a reliable transmission method, including wavelength of transmission, the type of modulation and polarization at which the signal will be transmitted. The range of wavelengths that can be used in transmitting the message is controlled by the following relationship:

Wavelength of transmission = (transmitter power × gain of transmitting antenna × gain of receiving antenna) / noise temperature of the receiving antenna

Because of the absence of information on the receiving antenna, one could transmit over a broad range of wavelengths depending on the parameters chosen in the above equation. Most conservative estimates restrict the wavelength range from 1 - 20 cm, which will make reception more likely for a civilization with modest technical capabilities. A civilization with more advanced capabilities has no reason not to detect a signal in this range. This gives us a reasonable constraint on the range of wavelengths that can be used for transmission, keeping the transmission system economical both technically and financially. This range can be increased with the availability of resources in order to broadcast over a broader frequency spectrum. Since the bandwidth of radio signals is quite large, it is highly improbable that the receiving antenna will be tuned at the particular frequency at which the message had been sent. One approach to this problem takes advantage of the common frequencies observed in nature within the radio range. A popular choice is the 21 cm emission line of interstellar neutral hydrogen [23, 25]. Using this frequency for communication in the galactic plane is noisy due to the background of interstellar 21 cm hydrogen; however, narrow-band signals near the 21 cm line that rapidly vary in brightness could still stand out due to the Doppler spread of hydrogen in the galaxy. Another close frequency based on universal constants is $21 cm/\pi = 6.72$ cm. Other combinations of fundamental constants for frequencies at which to transmit await further exploration [26].

Another important parameter is the type of modulation to be used in transmission. There are several modulation options available but frequency modulation (FM) is the simplest and most widely used one. Pulse modulation is another option and one that can be detected easier. A third option would take advantage of polarization modulation. Natural signals are typically randomly polarized, so any polarization with a specific pattern can convey the artificial nature of our message. (Natural signals that are polarized do not vary rapidly or periodically, and so they should be distinguishable from an artificially polarized signal.)

A METI protocol should include a set of transmission techniques and guidelines for making a broadcast. Ideally, a dedicated beacon would be established for conducting regular broadcasts, but this is a much longer-term ambition that will require significant international investment and cooperation. For the time being, there are a few antennas that have the capability to transmit the message at planetary distances at any location within our galaxy [23]:

Name of the telescope (disk diameter, antenna power, wavelength transmitted)

1. Arecibo, Puerto Rico (300 m; 1000 kW; 12.5 cm)

2. Solar System Planetary Radar in Goldstone, California (70 m; 480 kW; 3.5 cm)

    3. Planetary Radar near Evpatoria, Crimea (70 m; 150 kW; 6.0 cm)

LASER transmission is yet another very powerful means of communication. A powerful LASER, because of its narrow beaming angle, can outshine a star at a distance of a few light years. This makes it easy to be identified as an artificial signal from the normal background. A narrow beam LASER broadcast must be aimed at a particular target, though, while an omnidirectional fan beam radio transmitter can broadcast across the entire sky.

**5. Public Evaluation of the Protocol.**

Once a METI protocol has been developed, we suggest to use the protocol as a framework by which to test communication across human cultural boundaries and engage the public in thinking about METI. Although the idea of constructing a message free of anthropocentrism and testing it on humans sounds paradoxical, it is the most reasonable method of testing given the absence of any other species capable of building radio telescopes (which SETI defines as the lower limit of intelligence). Even if the METI protocol is never consistently used for messaging extraterrestrials, it will still provide a unique educational tool for science outreach to students and the public.

    Through an interactive website, we can allow users to create their own messages that conform to the protocol. Whatever the constraints of the protocol may be, there will nonetheless be some degree of freedom by which a user can decide what message they think is important. Messages submitted through the website can then be retrieved by other users, who then attempt to decrypt the message to see if the communication attempt was successful. Using a tool such as

this, we will be able to simultaneously 1) test the METI protocol and see where improvement is needed, and 2) collect the messages to extraterrestrials that at least some portion of the public would like to send. On this second objective, even if the messages are never sent, they will almost certainly provide novel information about the values of our culture.

The effectiveness of the protocol can be tested through international educational efforts by sending messages across cultural boundaries. For example, students in the United States may construct a set of messages according to the METI protocol, which are then exchanged with students in China. The differences in cultures between the respective student groups will likely be reflected in the message composition, so not all messages may be successfully communicated. In this case, students in the United States may discover that their particular ideologies that they believed to be obvious were somewhat lost in translation when deciphered by their international colleagues. Extension of these efforts to a diverse sample of human populations will help refine the protocol to be more universally decipherable. Evaluating the protocol across cultural boundaries on Earth will help refine the protocol while also fostering educational collaboration between human groups on Earth. Although this process will not remove all anthropocentric bias, it will help identify some cultural biases in the protocol.

It is certainly possible that messages collected through an interactive website may be suitable for broadcast, and there even may be several user submitted messages that withstand a wide array of cross-cultural scrutiny. Nevertheless, it seems more likely that, at least prior to contact, our messages to extraterrestrials will be based on observable features of the physical universe and contain mostly mathematics and science [24]. The call for public input may help improve the effectiveness of the protocol itself, but it will also actively engage the public in thinking about METI. After all, the first inclination of many users will be to communicate something deeply religious or humanistic, but they may find that their encoded message fails to effectively communicate to a different human culture. Involving the public in message construction in this way will increase awareness of cultural assumptions that are implicit in every society. This activity will also increase public understanding of Earth and humans in the context of astronomical spatial and temporal scales, which will hopefully inspire people to reduce catastrophic risk [27] and prepare for the future.

## 6. Conclusion.

A METI protocol is needed in order for a unified and international effort to be made in messaging extraterrestrials. By carefully constructing a framework by which to write and send messages, we will optimize the quality of messages as they are broadcast and increase the probability that we are understood. Additionally, the release of a protocol on an interactive public website will allow users ranging from scientists to students to test the protocol and construct sample messages. These encoded messages can then be tested across human cultural boundaries, which will provide a unique outreach opportunity to engage the public in thinking about METI. It is often said that SETI is a search for ourselves, and as we develop a message that we would send to unknown listeners, we will come to an even deeper appreciation of our diversity as humans.

**Acknowledgments**

We thank Carl Devito, Michael Busch, and Seth Baum for insightful comments that helped to improve this paper. Lingling Wu contributed to an earlier draft of this paper that was developed at the 2009 Astrobiology Graduate Conference (AbGradCon) Research Focus Group in Seattle, Washington.